# Discovering Melting Temperature Prediction Models of Inorganic Solids by Combining Supervised and Unsupervised Learning

Vahe Gharakhanyan[1,2], Luke J. Wirth[3], Jose A. Garrido Torres[4], Ethan Eisenberg[4], Ting Wang[4], Dallas R. Trinkle[3], Snigdhansu Chatterjee[5], and Alexander Urban[2,4,*]

[1]*Department of Applied Physics and Applied Mathematics, Columbia University, New York, NY 10027.*
[2]*Columbia Electrochemical Energy Center, Columbia University, New York, NY 10027.*
[3]*Department of Materials Science and Engineering, University of Illinois Urbana-Champaign, Urbana, IL 61801.*
[4]*Department of Chemical Engineering, Columbia University, New York, NY 10027.*
[5]*School of Statistics, University of Minnesota, Minneapolis, MN 55455.*
*\*a.urban@columbia.edu*

**Abstract**

The melting temperature is important for materials design because of its relationship with thermal stability, synthesis, and processing conditions. Current empirical and computational melting point estimation techniques are limited in scope, computational feasibility, or interpretability. We report the development of a machine learning methodology for predicting melting temperatures of binary ionic solid materials. We evaluated different machine-learning models trained on a data set of the melting points of 476 non-metallic crystalline binary compounds, using materials embeddings constructed from elemental properties and density-functional theory calculations as model inputs. A direct supervised-learning approach yields a mean absolute error of around 180 K but suffers from low interpretability. We find that the fidelity of predictions can further be improved by introducing an additional unsupervised-learning step that first classifies the materials before the melting-point regression. Not only does this two-step model exhibit improved accuracy, but the approach also provides a level of interpretability with insights into feature importance and different types of melting that depend on the specific atomic bonding inside a material. Motivated by this finding, we used a symbolic learning approach to find interpretable physical models for the melting temperature, which recovered the best-performing features from both prior models and provided additional interpretability.

## 1. Introduction

The compound melting temperature is an essential thermodynamic quantity for many processes, for example, for developing electrolytes for the electrolytic recycling of metals from spent electronics,[1–15] and for smelting processes to extract a base metal from its ore.[16–19] The melting point can be measured directly or obtained from phase diagrams when available for the chemical compositions in question, which is often not the case for non-metallic inorganic materials with high melting temperatures. For a large portion of inorganic chemical space, currently, no well-curated non-commercial public melting-temperature databases exist, to the author's knowledge.

Motivated by the challenges in the experimental measurement of melting temperatures, various theoretical and computational methods for estimating melting temperatures have been proposed. We distinguish between four general techniques of attaining melting temperatures: a) pure theoretical approaches based on theories that apply to specific classes of materials,[20–25] b) direct molecular dynamics (MD) simulations of melting, either based on first-principles calculations[26–35] or empirical interatomic potentials,[36–38] c) thermodynamic models based on the CALPHAD (CALculation of PHAse Diagrams) method,[39,40] and d) machine learning (ML) models for the direct prediction of melting temperatures without simulation.[41–46]

Theories describing melting have been mostly limited to metals and metallic alloys. The first such theory, the Lindemann melting criterion, proposed by Frederick Lindemann in 1910,[47] relates melting to the displacement of atoms from their equilibrium lattice sites: as temperature rises, the average amplitude of thermal vibrations in the lattice increases, and melting begins when the vibrations become severe enough that atoms collide with surrounding atoms. Lindemann used a closed-form equation to connect the melting point to either the Debye frequency or the Debye temperature. This criterion has been demonstrated to apply to single-component metal and metalloid systems,[20] but when the system's complexity increases, such as in the case of high-entropy alloys, modifications to this melting rule are required.[48–52] The Lindemann criterion does not generally apply to non-metallic compounds.

Direct molecular dynamics (MD) melting simulations are challenging with first-principles methods, such as density-functional theory (DFT),[53,54] because of the long timescales required at temperatures close to the melting point. However, various accelerated MD-based first-principles techniques have been proposed, such as the large-size coexistence method[37,55–57] and the fast-heating method.[58] DFT-based schemes have also been used to calculate the Gibbs/Helmholtz free energy values of solid and liquid phases and then solve for the melting temperature by equating these thermodynamic quantities for separate phases.[59,60] Despite such acceleration techniques, first-principles approaches are computationally demanding and limited to comparatively small simulation cells and timescales, reducing the accuracy of melting-temperature predictions. The computational cost of MD simulations is far lower with interatomic potentials, but accurate potentials that can describe both the crystal and the melt of an inorganic material are often unavailable.[61]

Thermodynamic models based on the CALPHAD approach can yield highly accurate phase diagrams, including melting transitions.[62,63] Unfortunately, few CALPHAD models of inorganic materials have been published, so model availability is a limiting factor.

The methodologies reviewed above are restricted to certain compound classes and are constrained by method accuracies and a lack of experimental data. These limitations motivated the development of techniques that combine ML models with physics-based models for melting-point prediction. Seko *et al.* demonstrated that incorporating DFT features into element-based materials fingerprints improves the predictive power of ML models for predicting melting temperatures of binary and single-component solid materials.[41] The authors found that support vector regression led to the best melting-temperature prediction model. Guan and Viswanathan developed an artificial neural network with Bayesian optimization for hyperparameter tuning that could successfully learn the melting temperatures of alloy systems using melting data from CALPHAD.[42] Hong and coworkers demonstrated melting temperature prediction for an extensive data set of more than 9,300 materials utilizing a graph neural network for materials embedding from compositions and a residual neural network for melting temperature prediction.[64–66] The data set used in their work mostly contained compounds with melting temperatures below 2,000 K (about 90% of the data set). As seen from the reviewed examples, the prior work on combining ML with first principles for melting temperature prediction was generally limited to certain types of materials and/or narrow ranges of melting temperatures.

In the present study, we address the limitations of melting temperature models through a combination of supervised and unsupervised learning. Our focus lies on binary inorganic solids with melting temperatures up to 4,000 °C because of the tremendous technological relevance[9,12] of this compound space. We particularly focused on binary inorganic materials because of the lack of experimental data and models.

However, we made sure that our featurization methods are transferable to more complex compositions such that, given the availability of an extended data set of more complex materials, new ML models can be fitted without the need to revise the materials embedding method. We considered additional unsupervised learning tasks to study the physics behind melting, motivated by the observation that melting is directly related to the bonding types within a material, and different bond-related features may affect melting differently (Lindemann criterion[48]). To this end, we employed unsupervised learning to partition our data set into separate groups based on bonding within materials. To learn more about the underlying physics of melting and approach interpretability, we used symbolic regression[67,68] to learn a closed-form equation for melting.

## 2. Methods

### 2.1. Model development

To develop a machine learning model to predict the melting temperatures of non-metallic, inorganic solids, we followed an approach consisting of the following five steps: (1) *data collection*, where we compiled a data set of experimentally measured melting temperatures from various sources; (2) *featurization*, where we explored a range of features and used domain knowledge to select the most appropriate ones for representing the materials; (3) *model selection and training/construction*, where we evaluated several machine learning models on the featurized data to identify the best-performing ones; (4) *model evaluation*, where we tested the performance of the best model on an independent data set to ensure generalizability; and (5) *result interpretation*, where we constructed symbolic regression models to understand the underlying relationships better. Detailed descriptions of each step are provided in the following subsections.

#### 2.1.1. Melting-temperature data

We compiled a data set of 476 binary ionic compound melting temperatures spanning a temperature range of 0–4,000 °C. **Figure S1** provides violin plots that show the distributions of these melting points over the various partitions of our data set. The experimental melting temperature data was primarily obtained from two CRC handbooks.[69,70] Compounds in our data set are exclusively of the metal-nonmetal, metal-metalloid, and metalloid-nonmetal element combinations, with metal-metal (alloys with predominantly metallic bonding), metalloid-metalloid and nonmetal-nonmetal (with mostly covalent bonding) compound types excluded. We imposed the constraint on compound types since we are primarily interested in ionic compounds and minerals. Note that the records of compounds with high melting points (>2,000 °C) often report ranges of temperatures rather than single points. Also, melting temperature data from different sources occasionally had conflicting values. In these cases, we include the average reported melting temperature as one single point in our data set for a given compound.

We partitioned our melting temperature data set into training and test sets for ML model training. The first set was used to build models, while the second was only used to evaluate them. We included those materials in the test set whose elastic moduli were not available in the Materials Project database[89,90] and were calculated by ourselves (see **Section 2.4** for details). This makes the test set performance especially meaningful since the model will, in practice, be primarily evaluated for compounds that have not previously been characterized, and thus, the elastic moduli values will have to be manually calculated. This resulted in a 420/56 (88%/12%) training/test split.

**Figure S2** illustrates the elemental distributions within the training and test sets. In terms of the statistical distribution of element types, the test set is representative of the training set. It is apparent from the melting temperature distributions that the test set is biased to lower melting temperatures. We consider this acceptable since the experimental values for high melting temperatures are generally subject to significant uncertainties.

Table I. Compound and elemental features, their symbols, and units.

| Property of | Feature | Symbol | Unit |
|---|---|---|---|
| Compound | Cohesive energy | $E_{\text{coh}}$ | eV/atom |
| | Formation energy | $H$ | eV/atom |
| | Bulk modulus | $K$ | GPa |
| | Shear modulus | $G$ | GPa |
| | Density | $\rho$ | g/cm$^3$ |
| | Bond ionic character | %IC | - |
| Element | Molar mass | $M$ | g/mol |
| | Atomic number | $Z$ | - |
| | Atomic radius | $R$ | Å |
| | Molar volume | $V_{\text{m}}$ | m$^3$/mol |
| | Pauling electronegativity | EN | - |
| | Periodic row number | $N_{\text{R}}$ | - |
| | Periodic group number | $N_{\text{G}}$ | - |
| | Elemental melting temperature | $T_{\text{m}}$ | K |

*2.1.2. Feature construction and selection*

We based our materials representation on a combination of *compound features*, i.e., properties of a given compound, and *element-based features*, i.e., features built from elemental properties using different statistical moments and averaging approaches, as shown in **Table I**.

Except for the bond ionic character, the compound features were obtained from DFT calculations extracted from the Materials Project,[71] using the most stable entry for the given composition, i.e., the entry with energy on or closest to the lower convex hull of formation energies. We performed DFT calculations of the bulk and shear moduli (see **Section 2.4** for details) for those compounds for which elastic moduli values were unavailable in the Materials Project (or were only available as ML predictions)[72] and included these compounds in the independent test set. The bond ionic character %IC was originally suggested by Pauling[73] as a measure of the percentage of ionicity in the bonds between two atoms A and B and is constructed from elemental electronegativities as $\%\text{IC} = \left(1 - \exp\left[\frac{(\text{EN}_A - \text{EN}_B)^2}{4}\right]\right) \cdot 100\%$, where $\text{EN}_i$ is the electronegativity of element $i$.

To obtain compound material features from elemental properties, we employed five statistical moments and averaging approaches, either accounting for stoichiometry/composition or only based on the combination of elements: the arithmetic average, standard deviation, harmonic average, quadratic average, and geometric average (see **Table II**). These moment and averaging methods were chosen to ensure that the features are symmetric with respect to the permutation of atomic species. The features additionally generalize without modifications to simpler unary systems and more complex systems (ternary, quaternary solids, or higher-order compositions).

Overall, we obtained 86 features (6 compound features: 5 DFT features from the Materials Project and the bond ionic character, and 80 composed features from 8 elemental features with 5 statistical moment and averaging methods and both composition-based and composition-agnostic weighting). This means our materials fingerprint encodes elemental properties, compositions, and compound-specific physical properties. We note that the inclusion of DFT-based features makes it possible to distinguish between polymorphs, such as, for instance, rutile and anatase titanium dioxide.

**Table II.** Statistical moments and averaging methods for elemental feature Z and compound $A_xB_y$.

| Method | Notations for composition-based and composition-agnostic features, respectively | Equations for composition-based features (for composition-agnostic features, $x = y = 1$) |
|---|---|---|
| Arithmetic average | $\langle Z \rangle_{\text{aw}}$ and $\langle Z \rangle_{\text{a}}$ | $\dfrac{xZ_A + yZ_B}{x + y}$ |
| Standard deviation | $\langle Z \rangle_{\text{sw}}$ and $\langle Z \rangle_{\text{s}}$ | $\dfrac{|xZ_A - yZ_B|}{x + y}$ |
| Harmonic average | $\langle Z \rangle_{\text{hw}}$ and $\langle Z \rangle_{\text{h}}$ | $\dfrac{x + y}{\dfrac{x}{Z_A} + \dfrac{y}{Z_B}}$ |
| Quadratic average | $\langle Z \rangle_{\text{qw}}$ and $\langle Z \rangle_{\text{q}}$ | $\sqrt{\dfrac{xZ_A^2 + yZ_B^2}{x + y}}$ |
| Geometric average | $\langle Z \rangle_{\text{gw}}$ and $\langle Z \rangle_{\text{g}}$ | $\sqrt[x+y]{Z_A^x Z_B^y}$ |

Applying these different averaging methods to our set of eight elemental features led to the creation of expectedly correlated features, as shown by a Pearson correlation coefficient analysis in **Figure S3**. We provided our model with all of these features to investigate which averaging method would perform best. Furthermore, the best-performing models were mostly ensemble tree-based, and based on our experience, even minor non-collinearity in features can result in better ensemble tree-based models.

*2.1.3. Supervised-learning model selection and evaluation*

We considered different regression models: (a) a linear regression model for benchmark purposes, as implemented in *scikit-learn* Python package,[74] (b) Gaussian process regression, kernel ridge regression, and support vector regression models (*scikit-learn*), (c) a vanilla light gradient boosting machine (LGBM) and a standard LGBM regressor, as implemented in *lightGBM*,[75] and (d) a random forest (RF) model, LGBM RF, as implemented in the *xgboost*[76] and *lightGBM* Python packages, respectively.

The performance of the models was measured by several error metrics: the mean absolute error (MAE), the root-mean-square error (RMSE), the mean absolute percentage error (MAPE), and the coefficient of determination ($R^2$). These metrics were computed for the training set with five-fold (5-fold CV), leave-one-out cross-validations (LOO-CV), and the test set. We constructed a hyperparameter grid for all the models, and hyperparameters were tuned with grid[47] and random search[77] techniques with five-fold cross-validation, as implemented in *scikit-learn*. Although there are too many parameters to carry out an exhaustive grid search easily, each parameter was manually adjusted, and the parameters that reduced the (average) RMSE over the five-fold cross-validation process were retained.

We tested several possibilities as the model output, including direct melting temperatures (in both K and °C units), logarithmic (to base 10) melting temperature (in K), and melting temperature from Vegard's law as $T_{\text{m,Vegard}} = T_{\text{m}} - \sum_i^{\text{Elements}} x_i T_{\text{m}}^{(i)}$, where $T_{\text{m,Vegard}}$, $T_{\text{m}}$, $T_{\text{m}}^{(i)}$, and $x_i$ are the melting temperature from Vegard's law, the melting temperature of the compound, as well as the melting temperature and atomic fraction of element $i$, respectively. We also considered different data normalization techniques on the input features, specifically, standardization and min-max scaling as implemented in *scikit-learn*. Models with the best performance often resulted from training using standardized features as inputs and the logarithm of the melting temperatures as outputs.

*2.1.4. Unsupervised-learning model construction*

For the clustering of the materials into different compound groups, we performed k-means clustering[78] using a subset of hand-picked features related to the bond ionicity (bond ionic character and weighted arithmetic average and weighted standard deviation of electronegativity, periodic row, and periodic group numbers). The optimal number of clusters was determined by simultaneous Silhouette score[79] and Calinski-Harabasz score[80] analyses when considering cases of up to ten clusters. Silhouette analysis shows the separation distance between the resulting clusters (how similar a data point is to its own cluster compared to other clusters using Euclidean distance). The Calinski-Harabasz score is the ratio of the sum of between-clusters dispersion and inter-cluster dispersion for all clusters and measures the separation between clusters based on Euclidean distances and the compactness of data points within each cluster. The Silhouette score and Calinski-Harabasz score analyses are presented in **Figures S4, S5, and S6**, along with the clustering results projected on a two-dimensional space with principal component analysis (PCA) in **Figure S7**.[81]

With the combined unsupervised/supervised model, predicting the melting temperature of a material from the test set involves two steps. Firstly, the cluster to which the test point belongs is identified using the (unsupervised) k-means model, and then the regression model for the selected cluster is used to make the temperature prediction.

*2.2. Symbolic regression model construction and validation*

We employed the variable-selection assisted sure independence screening and sparsifying operator (VS-SISSO)[68] to construct physically interpretable descriptors of the melting temperatures. VS-SISSO composes derived features from a set of input physical features using a set of operators. The present work uses $\{x_1 + x_2, x_1 - x_2, |x_1 - x_2|, x_1 \times x_2, x_1 \div x_2, \exp(x), \exp(-x), \ln(x), x^{-1}, x^2, x^3, \sqrt{x}, \sqrt[3]{x}\}$ set of operators, where $x_i$ can be either an input or a previously derived feature. Models take the form $T_m = c_0 + c_1 X_1 + \cdots + c_D X_D$, where $c_i$ are linearly optimized coefficients that minimize the training RMSE given a combination of $D$ derived features $X$. Each model has a dimensionality $D$, i.e., the number of derived features that the model depends on, and a feature complexity $f$, which is the maximal number of operations used within one derived feature.

The iterative variable selection of VS-SISSO (compared to regular SISSO)[67] allows for efficient screening of large feature spaces by considering subsets ($S$) of the entire feature space that combine new features ($S_a$) with those of the best-performing models ($S_b$). Here, model construction uses nine different sets of hyperparameters where $D$ and $f$ take on values of 1, 2, or 3. $S_a$ introduces four physical features at each step, $S$ has a maximal size of 16, and model construction is considered converged when no improvement in training RMSE is seen over 50 iterations. Up to 10,000 of the best models are output from each calculation, and for 1D, 2D, and 3D models, the SIS-selected subspace of derived features contains a maximum of 100,000, 10,000, and 1,000 items, respectively.

*2.3. SHAP analysis*

SHAP (SHapley Additive exPlanations) is a model interpretability method that uses Shapley values from game theory to estimate the predictive importance (SHAP score) of the features of a model.[82] Note that the SHAP method explains the ML model and not the data: SHAP importance is not a measure of the importance of a given feature according to the physics of a problem, but it instead shows how important a feature is to a given ML model.

We calculated the SHAP values to understand the predictions of the constructed models. The *shap* Python module[83] was used to estimate these values. SHAP values are generated for each model feature and have the same unit as the anticipated target (melting temperature or its derivatives). In our case, the best models resulted from training on logarithmic melting temperatures and, hence, the scale of the SHAP values. Furthermore, SHAP values are additive, meaning that adding all the SHAP values for a specific prediction plus a base value (the mean of the target melting temperature values of the data set) yields the model prediction value. For example, a SHAP value of 0.5 indicates that adding that particular feature value to which the SHAP value corresponds will increase the logarithmic melting temperature by 0.5 units from the base (mean) value.

We visualize SHAP analysis in the form of layered violin plots. The ten most important features are displayed in decreasing order of importance from top to bottom in these charts. The higher the importance, the greater the influence of that feature on the melting temperature prediction for the given model. For visualization purposes, for each SHAP value, the lighter colors (lower feature values) are closer to the horizontal line, while the darker colors (higher feature values) are closer to the edges of the layered violin plot.

*2.4. Elastic moduli computations with DFT*

A well-converged stress tensor is essential to derive correct elastic constants, such as bulk and shear moduli, from DFT. All DFT results reported in the present work were based on the projector augmented wave (PAW) method[84,85] as implemented in the Vienna Ab Initio Simulation Package (VASP).[26,86,87] We used the exchange-correlation functional by Perdew, Becke, and Ernzerhof (PBE).[88] The plane-wave energy cutoff was 700 eV, and the k-point density was 7,000 pra (per reciprocal atom). Energies converged to less than 0.01 meV. The atomic positions and lattice constants were relaxed until the residual forces became less than 0.001 eV/A. When the elastic tensor or the ionic-relaxation step failed to converge, the computation was restarted, but this time with new DFT parameters. As a result, the numerical parameters reported above were typical of many of our computations, but in certain circumstances, other parameters were employed (energy cutoff of 520 eV and/or k-point density of 1,000 pra).

To ensure compatibility with the data gathered from the Materials Project,[89,90] calculations of the moduli in this work take the Voigt–Reuss–Hill (VRH) average form[91] of the bulk modulus $K_{\text{VRH}}$ and the shear modulus $G_{\text{VRH}}$ calculated as

$$K_{\text{VRH}} = \frac{(K_{\text{V}} + K_{\text{R}})}{2} \quad \text{and} \quad G_{\text{VRH}} = \frac{(G_{\text{V}} + G_{\text{R}})}{2} \ ,$$

**Table III.** Error metrics for the best models for both the one-step approach and individual clusters in the two-step approach. We report LOO-CV and 5-fold CV scores for the training set. We report MAE, RMSE, MAPE, and $R^2$ metrics. All scores listed are the average metrics for models across ten different seeds. For MAE and RMSE, besides the average metric, standard deviations are also listed. For the two-step approach, we also report a metric evaluated on the combination of clusters 1 and 2 to facilitate the comparison with the one-step approach.

| Approach | Cluster | MAE [K] | RMSE [K] | MAPE [%] | $R^2$ |
|---|---|---|---|---|---|
| **Training set (LOO-CV)** | | | | | |
| One-step | – | 177±1 | 243±2 | 14 | 0.91 |
| Two-step | 1 | 169±2 | 246±4 | 17 | 0.89 |
| Two-step | 2 | 198±1 | 267±1 | 13 | 0.89 |
| Two-step | Combined | 184±2 | 257±3 | 15 | 0.89 |
| **Training set (5-fold CV)** | | | | | |
| One-step | – | 201±1 | 266±2 | 17 | 0.89 |
| Two-step | 1 | 189±3 | 264±4 | 19 | 0.87 |
| Two-step | 2 | 211±2 | 276±2 | 14 | 0.88 |
| Two-step | Combined | 201±2 | 271±3 | 17 | 0.88 |
| **Test set** | | | | | |
| One-step | – | 165±4 | 218±7 | 15 | 0.89 |
| Two-step | 1 | 136±5 | 169±6 | 15 | 0.92 |
| Two-step | 2 | 189±7 | 229±6 | 16 | 0.90 |
| Two-step | Combined | 156±4 | 194±4 | 15 | 0.92 |

where the subscripts V and R indicate the upper (Voigt) and lower (Reuss) bounds, respectively, defined by

$$9K_V = (c_{11} + c_{22} + c_{33}) + 2(c_{12} + c_{23} + c_{31}) ,$$
$$\frac{1}{K_R} = (s_{11} + s_{22} + s_{33}) + 2(s_{12} + s_{23} + s_{31}) ,$$
$$15G_V = (c_{11} + c_{22} + c_{33}) - (c_{12} + c_{23} + c_{31}) + 3(c_{44} + c_{55} + c_{66}) ,$$
$$\frac{15}{G_R} = 4(s_{11} + s_{22} + s_{33}) - 4(s_{12} + s_{23} + s_{31}) + 3(s_{44} + s_{55} + s_{66}) .$$

In the above equations, $c_{ij}$ and $s_{ij}$ are the elements of the stiffness and compliance tensors, respectively.

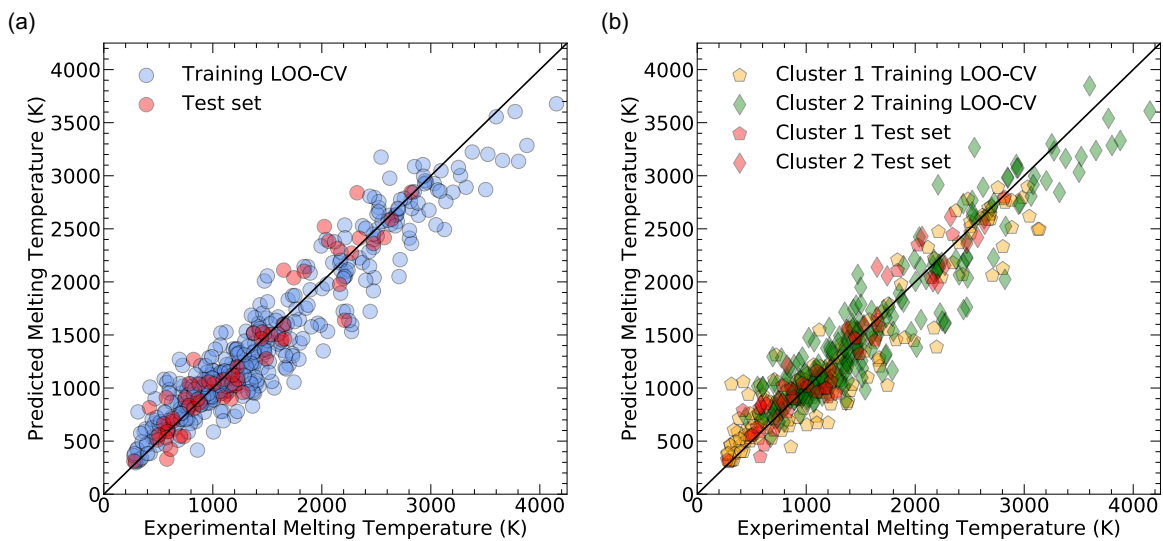

**Figure 1**. Correlation plots between the experimental and predicted melting temperatures for the best models trained on the **(a)** whole training data set and **(b)** training sets of clusters 1 and 2 separately. The training LOO-CV and test set labels represent ensemble averaged predictions over ten seeds. **Table III** shows the reported error metrics. The feature set used for training the model in (a) and for cluster 1 in (b) included compound features and both composition-based and composition-agnostic harmonic and quadratic average features of elemental properties. The feature set used for training the model for cluster 2 in (b) included compound features and both composition-based and composition-agnostic harmonic and arithmetic average features of elemental properties. The best model for the one-step approach in (a) resulted from the gradient-boosted tree regression, while random forest regression was the best model for clusters 1 and 2 in (b).

### 3. Results

#### 3.1. One-step approach: direct regression models

We first used direct supervised learning methods to construct models for melting temperature prediction by training on the entire training set. The best model resulted from an ensemble gradient-boosted tree regression (GBR) with a feature set that included compound properties and harmonic and quadratic (composition-based and composition-agnostic) averages of elemental properties (**Figure 1a**). We performed ensemble averages using ten different random seeds (statistics are presented in **Figure S8**), and the error metrics are presented in **Table III**. **Figure 1a** shows the correlation of the predicted melting temperatures from the best model of the one-step approach with the reference experimental values. This figure shows that the predictions strongly correlate with the actual values and are scattered around the perfect diagonal without a systematic error. **Figure S9** shows that the distribution of the errors for the test set is similar to that of the training set, further indicating that the model is not overfitted. An overview of the performance of other considered models is shown in **Table S1**.

    **Figure 2** presents feature importance through ensemble feature importance[92] and SHAP analysis.[82] Both feature importance methods identified the cohesive energy and elastic moduli features as the most important.

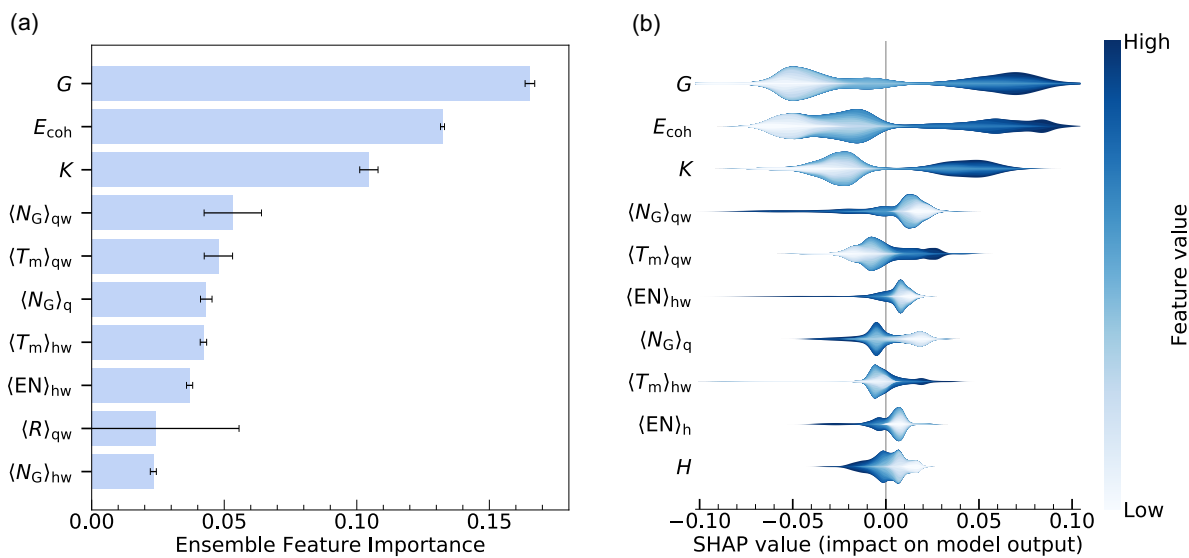

**Figure 2.** Feature importance for the one-step approach and the best ensemble model trained on the whole training data set computed using **(a)** the ensemble tree method and **(b)** the SHAP analysis. The features are ranked in order of their importance from top to bottom. Only the top ten most important features are present for both approaches. The ensemble tree importance analysis in (a) represents ensemble-averaged results over ten seeds, and the horizontal error bars indicate values of importance one standard deviation above and below their respective average values over ten different seeds. In (b), a positive/negative SHAP value shows that the feature increases/decreases the melting temperature prediction. For each SHAP value in (b), the lighter colors (lower feature values) are closer to the horizontal line, while the darker colors (higher feature values) are closer to the edges of the layered violin plot.

*3.2. Two-step approach: combined clustering and regression models*

The feature-importance analysis of **Figure 2** identified the shear modulus and cohesive energy as the two most important features for melting temperature prediction with a tree-based model. As reviewed in the introduction, melting occurs when the binding between atoms is partially overcome, so the importance of cohesive energy also agrees with intuition. The binding between atoms in a compound is due to different interactions, most importantly, covalent, ionic, and metallic bonding. The nature of the bonding determines, for example, how the cohesive energy changes with the coordination number. For instance, in ionic compounds, the cohesive energy is proportional to the coordination number, while it behaves as the square root of the coordination number in metals.

Motivated by these considerations, we investigated a second class of models involving an additional clustering step before the regression, in which materials were grouped based on features related to the bonding character. For this two-step approach, we first used an unsupervised-learning method to separate our data set into clusters and then fit supervised learning models within each cluster separately. For melting temperature prediction, firstly, the cluster a material belongs to is identified, and then the corresponding regression model is used for the melting point prediction.

Our clustering approach was based on k-means clustering.[78] Clustering was based on features that we intuitively associated with the bond ionicity (bond ionic character and weighted arithmetic average and
10

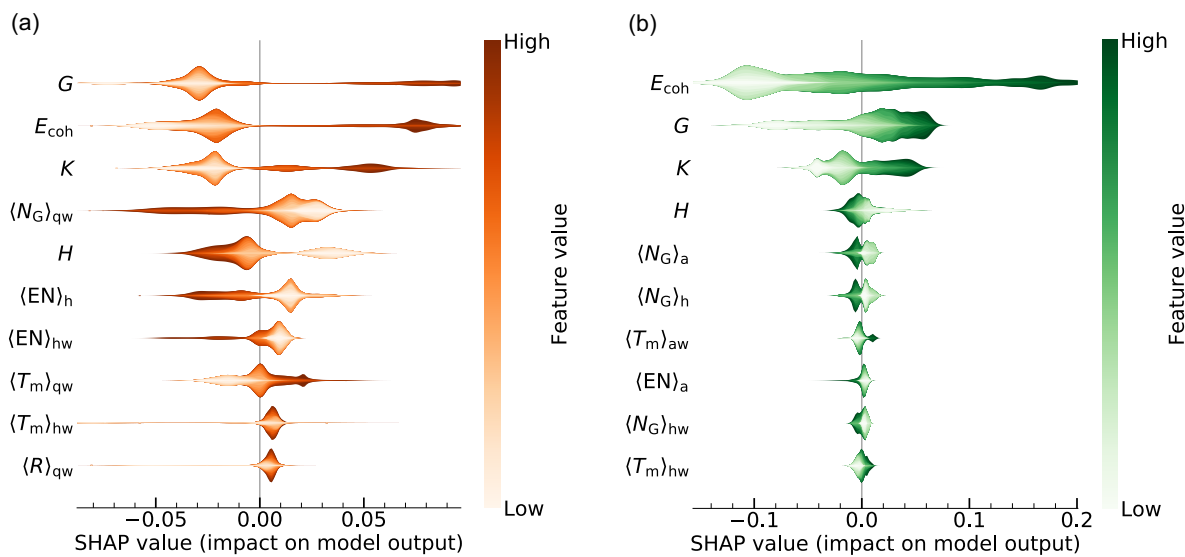

**Figure 3.** Feature importance analysis from the SHAP method for **(a)** cluster 1 and **(b)** cluster 2. The features are ranked in order of their importance from top to bottom. Only the top ten most important features are present for both clusters. A positive/negative SHAP value shows that the feature increases/decreases the melting temperature prediction. For each SHAP value, the lighter colors (lower feature values) are closer to the horizontal line, while the darker colors (higher feature values) are closer to the edges of the layered violin plot.

weighted standard deviation of electronegativity, periodic row, and periodic group numbers) to facilitate the interpretation of different melting behaviors based on bonding in our data set. The optimal number of clusters was chosen from simultaneous Calinski-Harabasz score[80] (**Figure S4**) and silhouette score[79] (**Figure S5 and Figure S6**) analyses, both of which resulted in an optimal number of two clusters. Cluster 1 and 2 contained 198 and 222 compounds, respectively. The clustering results projected on a two-dimensional space with principal component analysis (PCA) are presented in **Figure S7**.

Within each cluster, we fitted regression models for melting temperature prediction. The best models resulted from an ensemble random forest regression (RFR) for both clusters with feature sets that included compound properties and (both composition-based and composition-agnostic) harmonic and quadratic averages of elemental properties for cluster 1 and compound properties and (both composition-based and composition-agnostic) harmonic and arithmetic averages of elemental properties for cluster 2 (**Figure 1b**). We performed ensemble averages over ten different seeds (statistics are presented in **Figure S8**), and the error metrics are presented in **Table III**. An overview of the performance of other considered models is shown in **Table S1**.

We carried out a feature importance analysis through the ensemble feature importance (**Figure S10**) and SHAP method for both clusters (**Figure 3**).

To compare the results from the two-step approach to those from direct supervised learning of the previous section, we evaluated the overall error scores for the complete data set including both clusters (**Table III** and **Figure S11**).

**Table IV.** VS-SISSO-generated models for melting temperature for the one- and two-step approaches with the lowest average RMSE scores of the training and test sets. Two models are provided for the entire training set and cluster 2 cases since providing SISSO with the different types of averages yields slightly better performance. For cluster 1, the best models contain only the arithmetic averages. Here, we report RMSE and $R^2$ scores for the training and test sets.

| Approach | Cluster | $T_m$ [K] equation/model from VS-SISSO | Training set | | Test set | |
|---|---|---|---|---|---|---|
| | | | RMSE [K] | $R^2$ | RMSE [K] | $R^2$ |
| One-step | – | $389 + 56(\sqrt[3]{G} \cdot E_{\text{coh}} \cdot \langle R \rangle_a)$ | 368 | 0.78 | 337 | 0.75 |
| | – | $282 + 61(\ln(G) \cdot E_{\text{coh}} \cdot \langle R \rangle_{\text{qw}})$ | 342 | 0.81 | 310 | 0.79 |
| Two-step | 1 | $276 + 58(\sqrt[3]{G} \cdot E_{\text{coh}} \cdot \langle R \rangle_a)$ | 358 | 0.76 | 324 | 0.69 |
| | 2 | $730 + 129 \left( \frac{E_{\text{coh}}^2}{\sqrt{\langle N_G \rangle_a}} \right)$ | 318 | 0.84 | 287 | 0.84 |
| | 2 | $685 + 58 \left( \frac{E_{\text{coh}}^2}{\sqrt{\langle \text{EN} \rangle_{\text{hw}}}} \right)$ | 310 | 0.85 | 290 | 0.83 |

*3.3. Symbolic models from VS-SISSO*

While our model evaluation found tree ensemble methods to perform best for the melting temperature data set, these models are not readily interpretable. This makes it challenging to determine in which way the materials in the two clusters of the previous section behave differently when melting.

**Table IV** provides VS-SISSO-generated symbolic expressions for the melting temperature based on the entire data set as well as for the two clusters individually, giving qualitative insight into the melting physics that dominate within each one. Models selected for inclusion in **Table IV** have the lowest averaged training and test RMSEs of all generated 1D models with feature complexity $f = 3$ from the set of features that yields the best training RMSE.

## 4. Discussion

We constructed models for predicting the melting temperature of inorganic, non-metallic compounds using a direct one-step supervised learning approach and a two-step approach based on an additional unsupervised clustering step.

Fitting models without elastic moduli features yielded errors greater than 400 K, so we decided that incorporating and computing elastic features is necessary. We began our studies with a data set of about 600 compounds. For high melting temperature compounds that lacked elastic moduli data in the Materials Project database, we computed elastic moduli ourselves but experienced numerical and convergence errors in some cases. This led us to exclude these compounds from our data set, and as a consequence, very high melting temperatures are absent from the test set.

For the best-performing direct one-step regression model, the first and third most important features are the shear and bulk moduli, respectively. The elastic moduli relate to the resistance to elastic deformation when stress is applied to the compound. As such, these quantities are related to the bonding within the materials that determine how easily bonds within a material can be bent and eventually broken, and thus, is an intuitive measure for melting temperatures. Intuition tells us that soft, deformable materials have a lower melting temperature than hard, brittle materials. The second most important feature, cohesive energy, is the difference between the bulk energy and the sum of the energy of isolated atoms. It is also an intuitive feature since this energy needs to be partially overcome during melting.

While other top features, such as the statistical combinations of elemental periodic group number and elemental melting temperatures, are also important based on chemical intuition, our model cannot provide specific interpretable rules that relate these features with melting. This is a general limitation of black-box tree-based ML models since the decision criteria a model uses to generate predictions are often too complex to be interpreted directly. For this reason, we confirm our importance studies with SHAP analysis, which clearly shows the same ranking of feature importances for the five most important features for the model. For the first three most important features, elastic moduli and cohesive energy, SHAP analysis shows that high feature values tend to increase the predicted melting temperatures, while low feature values decrease the predicted melting temperature. This can also be seen in the correlation plot of the cohesive energy with the melting temperature (**Figure S12**) and the correlation plot of the cohesive energy with the elastic moduli (**Figure S13**). Higher cohesive energies result in higher melting temperatures.

Results from the two-step approach provide further insights. The distribution of elements within the two clusters (**Figure S14**) shows that cluster 1 contains most halides and oxides, while cluster 2 contains B-, C-, N-, and O-group compounds except oxides. Given the types of materials that are included in our data set and based on the elemental distributions, we conclude that cluster 1 incorporates more ionic compounds while cluster 2 comprises less ionic and more covalent solid materials. The melting temperature distribution of the two clusters (**Figure S1b**) shows that very-high (>3150 K) melting temperatures appear only in cluster 2. The compounds with extreme melting temperatures are mostly metal carbides known for their high melting temperatures resulting from a combination of inter- and intra-molecular covalent bonds that form covalent solid networks for this category of materials.[93,94]

For cluster 1, the ranking and relative importances of the first four features are similar to those found in the one-step approach for the entire data set. For cluster 2, though, we observe significant differences. By far, the most important feature for cluster 2 is cohesive energy related to the melting temperature, as described above. In the SHAP analysis of **Figure 3**, it is seen that very high cohesive energy values increase the predicted melting temperatures with a greater extent and, vice versa, very low cohesive energy values significantly decrease the predicted melting temperatures. While elastic moduli are the next important features for cluster 2, they have less impact on the model output when compared to the cohesive energy feature. Here, we conclude that cohesive energy across the entire range of values is crucial to melting point prediction for more covalent solid materials. We also see that, in general, other features are evenly split in terms of the feature value and SHAP value, meaning they enter the melting point prediction in more complex relationships.

**Table III** compares the predictive power of the one- and two-step approaches with tree ensemble methods. In terms of the training set LOO-CV scores, the MAE for cluster 2 is higher (by ~20 K) than that of the entire training set from the one-step approach, while the MAPE score is reduced because the fraction of very-high (>3150 K) melting temperature materials in cluster 2 is higher than that in cluster 1. The training set LOO-CV scores are very comparable among the one- and two-step approaches when computing the combined error metrics for clusters 1 and 2. The same can be noted for the training set 5-fold CV scores. Note that we generally observe lower RMSE and MAE error metrics from the test set because it is limited to a lower melting temperature range, so this does not indicate underfitting. The main differences are noticeable in the test set metrics. Specifically, the test set MAE and RMSE for cluster 1 are lower than that of the one-step approach by 29 K and 49 K, respectively. This also comes with the cost of increasing the aforementioned metric values for cluster 2. But overall, when comparing the whole test set to the combination of clusters 1 and 2, we see only a slight improvement in predictive power compared to the one-step approach (reduction in MAE of 9 K and RMSE of 24 K). This is not surprising since decision-tree-based models can internally perform a feature-based clustering so that a one-step tree-ensemble model can effectively replace the combined k-means clustering and regression steps.

However, this is not the case for other models, such as linear regression models, and here, the difference between the one-step and two-step approaches is much more significant. From **Table S1**, linear models (also with regularization) always result in MAE scores of greater than 240 K and RMSE scores greater than 340 K. These scores are at least 100 K higher than the respective best-performing models reported in **Table III** and are also significantly higher (by about 30 K) than what is attainable by the two-step approach when using the same linear models. Note that the linear models can be considered a benchmark of what is attainable with the simplest models, but they achieve poor metrics compared to the non-linear models we have evaluated in this work (**Table S1**). We also attempted model stacking but decided not to report more complex and less interpretable models with only slightly better performance.

Since clustering is beneficial for linear regression, we expected it also to facilitate symbolic regression with SISSO. For the models trained on the whole data set, the 11 best-performing 1D models with $f = 3$ with average train-test RMSEs of 326–382 K all involve combinations of the shear modulus ($G$), cohesive energy ($E_{\text{coh}}$), and the statistical combination of elemental atomic radii ($\langle R \rangle$). Model complexity can significantly affect feature selection; the top 15 1D and $f = 2$ models for both clusters all include $E_{\text{coh}}$ and the statistical combination of elemental electronegativities ($\langle EN \rangle$) while spanning a slightly higher range of average train-test RMSEs from 341–407 K. Larger models offer relatively limited gains in accuracy while becoming less parsimonious and therefore harder to interpret. For example, the top 3D model with $f = 3$ trained on the data from clusters 1 and 2 has training and test RMSE values of 312 and 299 K, respectively, comparable to the 1D values of 368 and 337 K. Another example, the best-performing 3D model with $f = 3$ trained on the cluster 2 takes the form $T_{\text{m}} = c_0 + c_1 \frac{E_{\text{coh}}^2}{\sqrt{\langle N_G \rangle}} + c_2 \frac{\langle \rho \rangle \cdot \langle EN \rangle \cdot \langle N_G \rangle}{\langle Z \rangle} + c_3 \frac{\exp(\langle EN \rangle)}{\sqrt[3]{\langle N_G \rangle}}$, and is significantly more complicated than the model shown in the bottom row of **Table IV**, but the averaged training and test RMSEs of this 3D model improve on those of the 1D one by only 22 K.

Comparing which physical features are commonly selected by SISSO across various model complexities can provide insight into which features are most essential to describing each cluster. The cohesive energy, $E_{\text{coh}}$, is present in all 1D models of **Table IV**, but it is often absent in cluster 1 models that perform nearly as well (e.g., with an average train and test RMSE of 377 K). On the other hand, the shear modulus, $G$, regularly appears in cluster 1 or one-step approach models but is never selected in a 1D cluster 2 model with $f = 2$. These patterns suggest that $G$ is particularly important for describing the melting temperature of cluster 1 compounds, and $E_{\text{coh}}$ is likewise the key for cluster 2. Additionally, as **Table IV** illustrates, cluster 2 models tend to have higher $c_0$ values. While this can lead to overestimating relatively low melting temperatures, cluster 2-only RMSEs are lower than in either case where cluster 1 compounds are present, which have intercepts near room temperature. We also investigated how exactly each feature enters the melting equation. In general, we noted that physical feature selection is more critical than operator choice for SISSO. Even though we noted that for ionic compounds, the cohesive energy is proportional to the coordination number, while it behaves as the square root of the coordination number in metals, as seen in **Figure S15**, there appears to be no distinct mathematical operator that leads to decreased test set RMSE metric when acted on $E_{\text{coh}}$ (although a few such as $E_{\text{coh}}^{-1}$ rarely appear in well-performing simple models).

Indeed, with a combined average test-set RMSE of 304 K, weighted by the number of materials in each cluster, the two-step SISSO model is slightly better than the model fitted to the entire data set with an RMSE of 310 K. However, the most interesting observation is that the data in cluster 2 can be represented well with symbolic models depending only on two features: the squared cohesive energy and the inverse square root of either the statistical combination of elemental periodic groups or the electronegativities (**Table IV**). The interchangeability of the periodic group and electronegativity is intuitive as the electronegativity strongly correlates with the group in the periodic table for cluster 2. The expression $T_\mathrm{m} \approx a + b \frac{E_\mathrm{coh}^2}{\sqrt{\langle \mathrm{EN} \rangle_\mathrm{hw}}}$ with $a = 685$ K and $b = 58$ K/eV$^2$ and the cohesive energy in eV can thus be used to obtain an estimate of the melting temperature of non-oxides and non-halides with melting points above 685 K. The symbolic regression studies also corroborate that the most important features are those picked up by feature importances and SHAP analysis of the tree ensemble models.

The linear model fitted on the entire 86-feature input results in the following metrics: for the training set LOO-CV – RMSE: 346 K, MAE: 259 K, for the training set 5-fold CV – RMSE: 378 K, MAE: 289 K, and for the test set – RMSE: 435 K, MAE: 353 K. Similar error metrics are observed for both cluster 1 and 2. Because of the computing requirements, we did not carry out LOO-CV or 5-fold CV studies for the VS-SISSO models. The test set RMSE scores for the linear models are higher than those from the VS-SISSO models with a difference of less than 100 K. This is a very notable advantage of using a symbolic learning model instead of a simple linear model, as even the least complex VS-SISSO models with small values of dimensionality (1D) and feature complexity (up to 3 considered in this work), like those shown in **Table IV**, can result in better fittings to the data and provide interpretability, albeit at an increased burden of computational complexity.

Although the present work is aimed at simplicity and interpretability, it is educative to compare our results to previously published models. Recently, Hong and coworkers published a deep learning model[43,65] based on a graph neural network for materials embedding from compositions and a residual neural network that is currently deployed on Microsoft Azure and the Research Computing facilities at Arizona State University. Their model is freely accessible and can be used to predict melting temperatures of compounds containing up to four elements (version 1) through API calls.[95] Unlike our data set, Hong's data contains entries from experiments and DFT-based simulations. *Hong* reports $R^2$ training score of 0.99, $R^2$ test score of 0.96, a training RMSE of 75 K, and a test RMSE of 138 K for their model for a data set containing melting temperatures of more than 9,300 materials. It took about 11 and 1.5 minutes to get predictions for our training (420 compounds) and test (56 compounds) sets, respectively, yielding the following metrics: for our training set – RMSE: 304 K, MAE: 200 K, MAPE: 20%, $R^2$: 0.85, and for our test set – RMSE: 404 K, MAE: 265 K, MAPE: 31%, $R^2$: 0.64. We note that our data set might contain materials that *Hong's* model was trained on (that might even have different melting temperature values tabulated from our data set), and we have made no effort to correct this bias. Our model thus outperforms *Hong's* model for both our training and test sets, even though *Hong's* model has overall lower reported error metrics and is generalizable and transferable to larger compositional materials spaces. While it is not an entirely fair comparison, since *Hong's* model is suitable for a broader range of materials and does not require features from DFT calculations, it is encouraging that our simple and computationally extremely efficient tree ensemble model can deliver state-of-the-art predictive performance.

## 5. Conclusion

We evaluated the predictive performance of different machine learning methods for melting temperature prediction of binary inorganic solid materials. We first considered a direct supervised-learning approach, finding that SHAP analysis identified the cohesive energy and elastic moduli as the most important features. We found that the fidelity of predictions can further be improved by introducing an additional unsupervised-learning step that first classifies the materials based on their bonding characteristics before melting-point regression. Not only does this two-step model exhibit improved accuracy, but the approach also provides additional insights into feature importance and different types of melting that depend on the degree of ionicity in the bonding inside a material. Finally, we employed symbolic learning to establish interpretable physical models for the melting temperature that highlighted the significance of the top-performing features from the previous approaches and gave more insight into the theory behind melting and led to a simple linear equation for estimating the melting temperature of non-oxides and non-halides.

**Data accessibility**

All experimental reference data, material fingerprints, and trained models have been posted on GitHub and are freely accessible at https://github.com/atomisticnet/MeLting.

**Code availability**

The ML methodology of the present work is implemented in Python. The provided code yields vectorized compound fingerprints and can generate predictive models based on a selected regression model, perform hyperparameter tuning, as implemented in *scikit-learn*, validate, and test the models. MeLting also uses the Python Materials Genomics (*pymatgen*) package[96] for materials fingerprint construction. The source code for this work has been published on GitHub at https://github.com/atomisticnet/MeLting and is freely accessible under the terms of the MIT license.

**Acknowledgments**

This work was supported by the National Science Foundation (NSF) under Grant No. DMR1940290 (Harnessing the Data Revolution, HDR). We acknowledge computing resources from Columbia University's Shared Research Computing Facility project, which is supported by NIH Research Facility Improvement Grant 1G20RR030893-01, and associated funds from the New York State Empire State Development, Division of Science Technology and Innovation (NYSTAR) Contract C090171, both awarded April 15, 2010. This work also made use of the Illinois Campus Cluster, a computing resource that is operated by the Illinois Campus Cluster Program (ICCP) in conjunction with the National Center for Supercomputing Applications (NCSA) and which is supported by funds from the University of Illinois Urbana-Champaign. The research of SC is partially supported by the National Science Foundation (NSF) under Grants No. 1939916 and No. 1939956.

# SUPPORTING INFORMATION

# Discovering Melting Temperature Prediction Models of Inorganic Solids by Combining Supervised and Unsupervised Learning

Vahe Gharakhanyan[1,2], Luke J. Wirth[3], Jose A. Garrido Torres[4], Ethan Eisenberg[4], Ting Wang[4], Dallas R. Trinkle[3], Snigdhansu Chatterjee[5], and Alexander Urban[2,4,*]

[1]*Department of Applied Physics and Applied Mathematics, Columbia University, New York, NY 10027.*
[2]*Columbia Electrochemical Energy Center, Columbia University, New York, NY 10027.*
[3]*Department of Materials Science and Engineering, University of Illinois Urbana-Champaign, Urbana, IL 61801.*
[4]*Department of Chemical Engineering, Columbia University, New York, NY 10027.*
[5]*School of Statistics, University of Minnesota, Minneapolis, MN 55455.*
*a.urban@columbia.edu

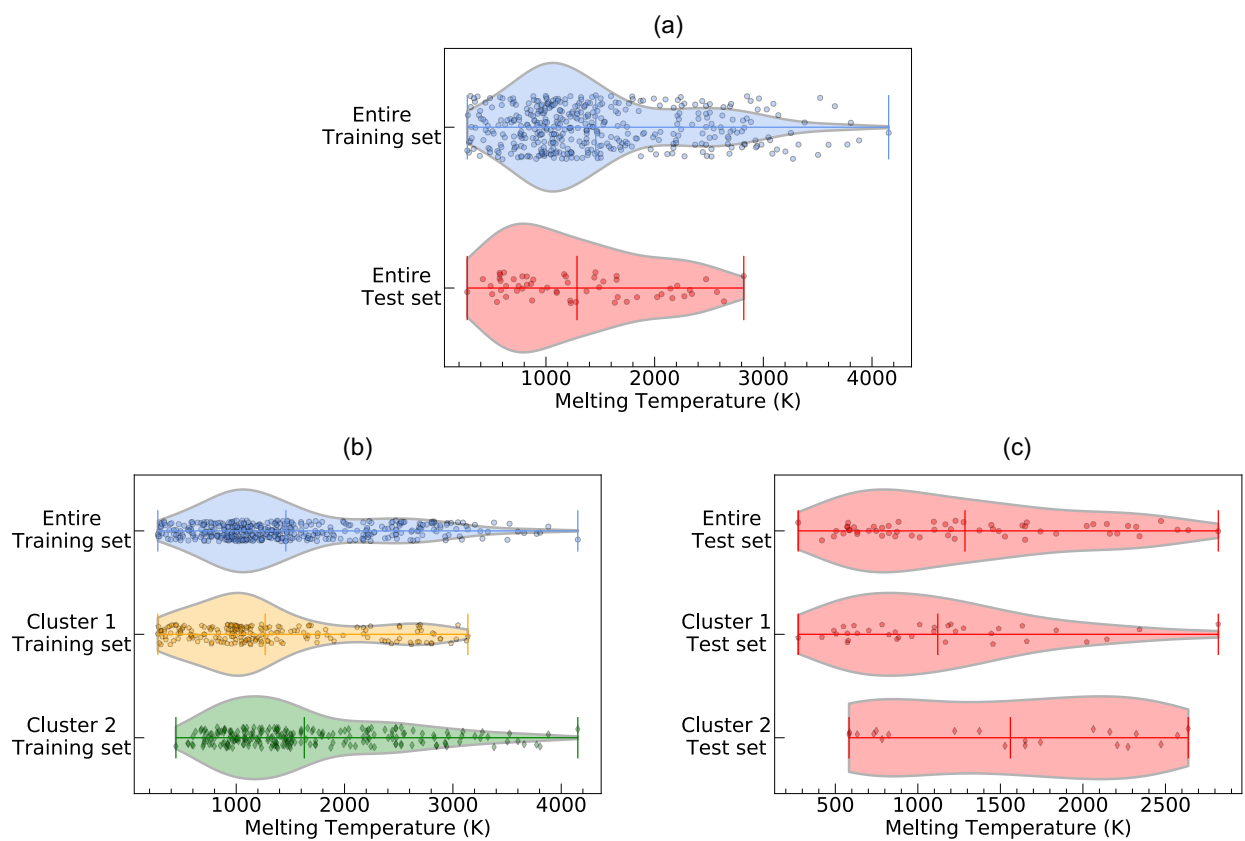

**Figure S1.** Comparison of the melting temperature distributions between (a) the entire training set and the entire test set, (b) the entire training set and the individual training sets of clusters 1 and 2, and (c) the entire test set and the individual test sets of clusters 1 and 2. For visualization purposes, different jitters are applied to each set. The violin plots show the density trace of the data and cover the whole data range. Each violin plot also shows the minimum, average, and maximum values of the set (in that order from left to right), with the vertical lines matching the data set color.

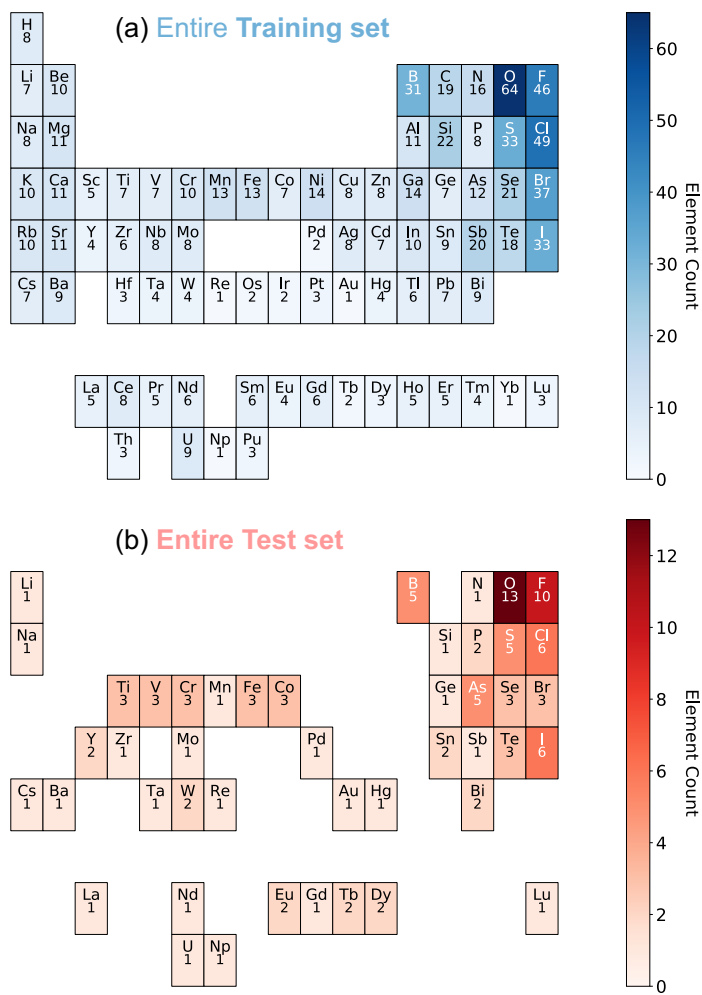

**Figure S2.** Elemental counts of the (a) training and (b) test sets of the entire data set. These periodic tables show the elements found in the data set and, below the element symbol – the number of element occurrences. Elements not found in the binary compounds in the given data set are omitted from the periodic table figure for the respective set. For instance, no hydrides are found in the test set, while eight are in the training set. The most occurring elements and their numbers of occurrence are colored in white.

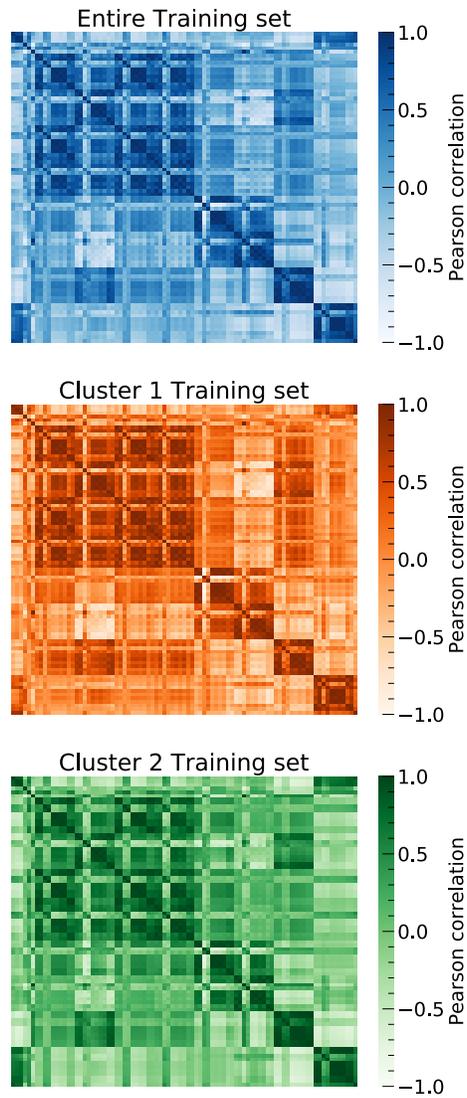

**Figure S3.** Pearson correlations between all 86 features for the entire training set and training sets of each cluster separately. The exact feature pairs are not shown for visualization purposes, but the general trends are more important. The ordering of all features is the same for all correlation plots.

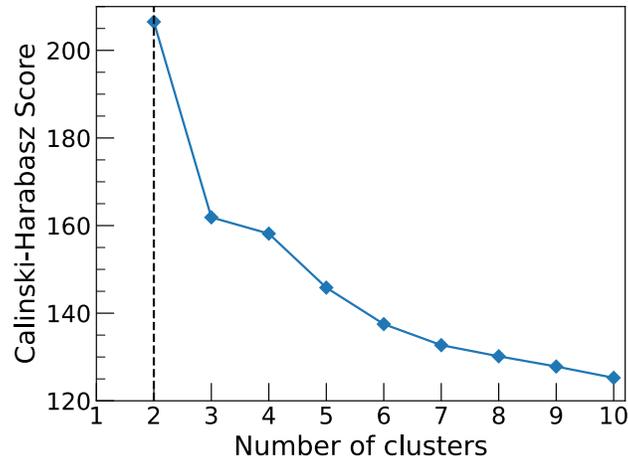

**Figure S4.** Calinski-Harabasz score versus the number of clusters for k-means clustering of the whole training set. We considered up to ten clusters in this work. Higher Calinski-Harabasz scores are desired. The highest score appears at two clusters and matches the result from the Silhouette score analysis.

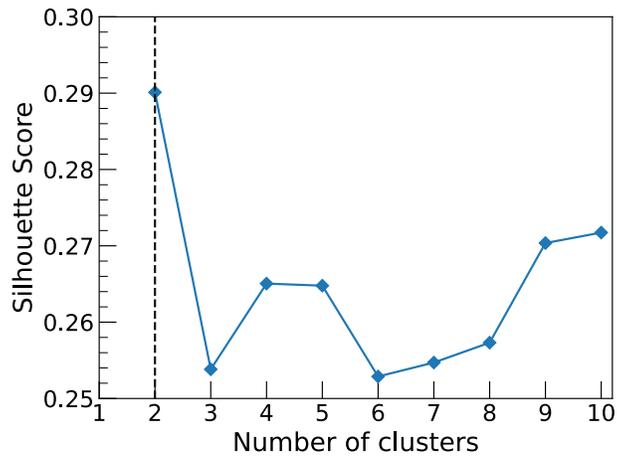

**Figure S5.** Silhouette score versus the number of clusters for k-means clustering of the whole training set. We considered up to ten clusters in this work. Higher Silhouette scores are desired. The highest score appears at two clusters.

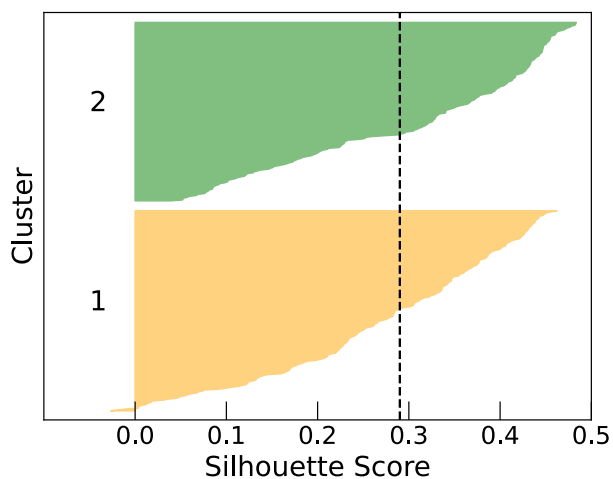

**Figure S6.** Silhouette score distribution for clusters 1 and 2. The vertical black line shows the average Silhouette score across both clusters. The color codes represent different clusters, and pictured are the silhouette score distributions ranked from the highest (top) to lowest (bottom) for each cluster. Cluster 1 has some negative Silhouette scores, but the average score of 0.29 is still high.

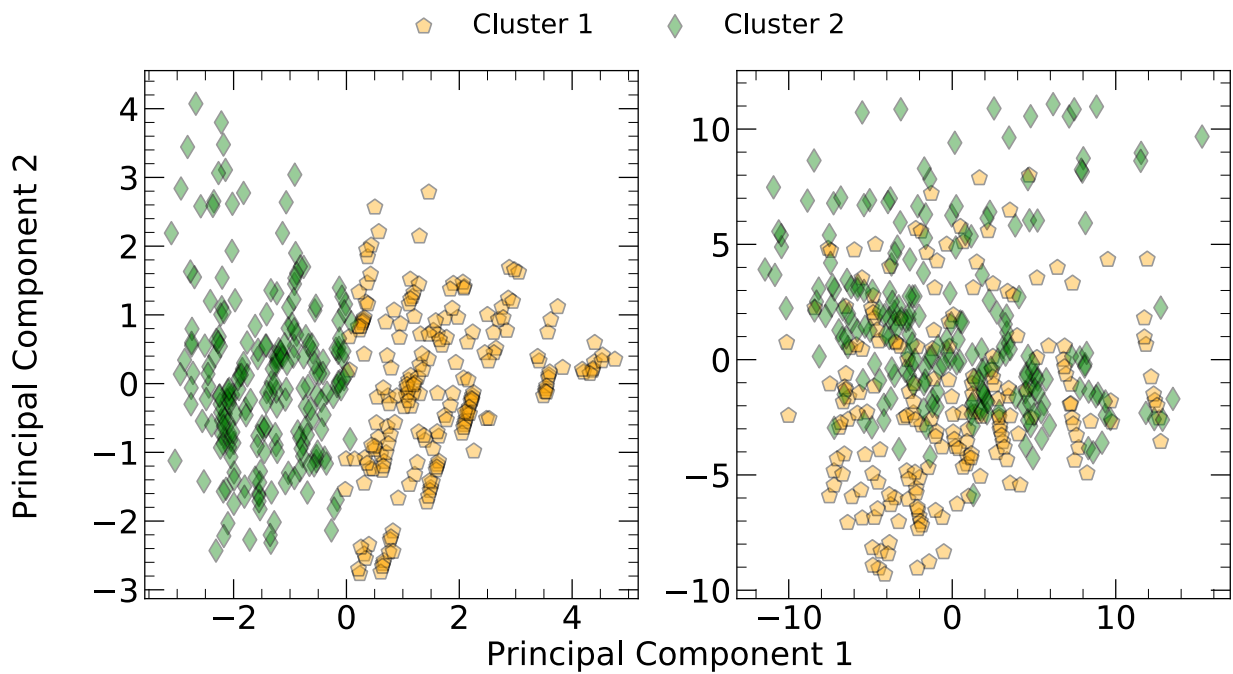

**Figure S7.** Distributions of the first two principal components of the entire training set by the cluster belongings. The left subfigure shows the principal components based on hand-picked bond-ionicity features only, and a separation of clusters is observed, while the right subfigure shows the principal components based on all features, which shows that the hand-picked features are not enough for the separation of the entire feature set. Hence, the difference in the scales of principal component values. This indicates that k-means clustering based on hand-picked bond-ionicity features provides a different separation than a naïve k-means clustering on the entire feature set.

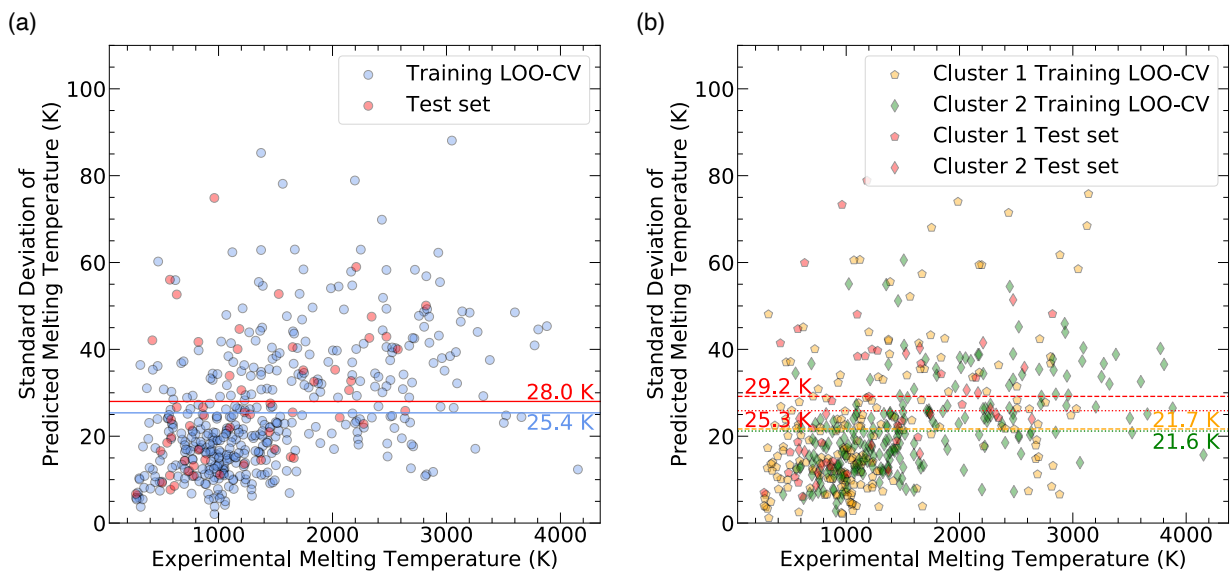

**Figure S8.** Standard deviations of the predicted melting temperatures over ten different seeds trained on the (a) entire training set and (b) training sets of clusters 1 and 2 separately versus the experimental melting temperatures. The best model for the one-step approach in (a) resulted from the gradient-boosted tree regression, while random forest regression was the best model for clusters 1 and 2 in (b). We can observe that the average standard deviation increases with the increase of melting temperatures, and thus the model confidence diminishes. The horizontal lines show the average standard deviations for the respective set. In (b), the dashed lines correspond to cluster 1, and the dotted lines correspond to cluster 2.

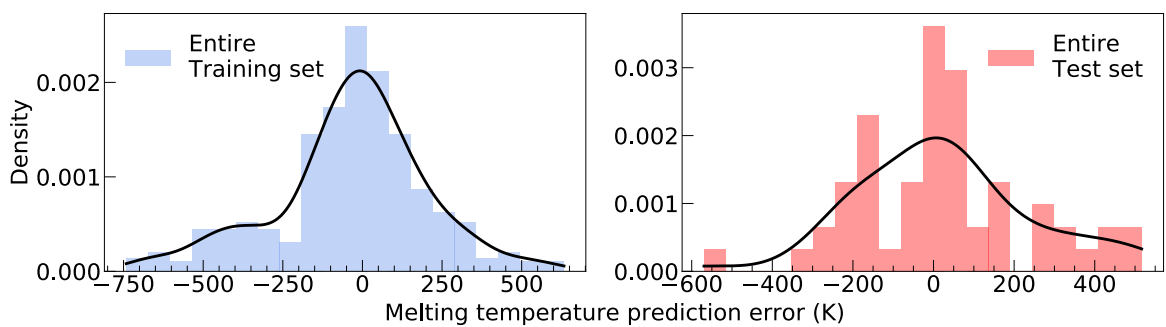

**Figure S9.** Distributions of the melting temperature prediction errors for the entire training and test sets from the one-step approach. For each subfigure, both a histogram and a density plot are shown. The distributions are normally distributed around the origin.

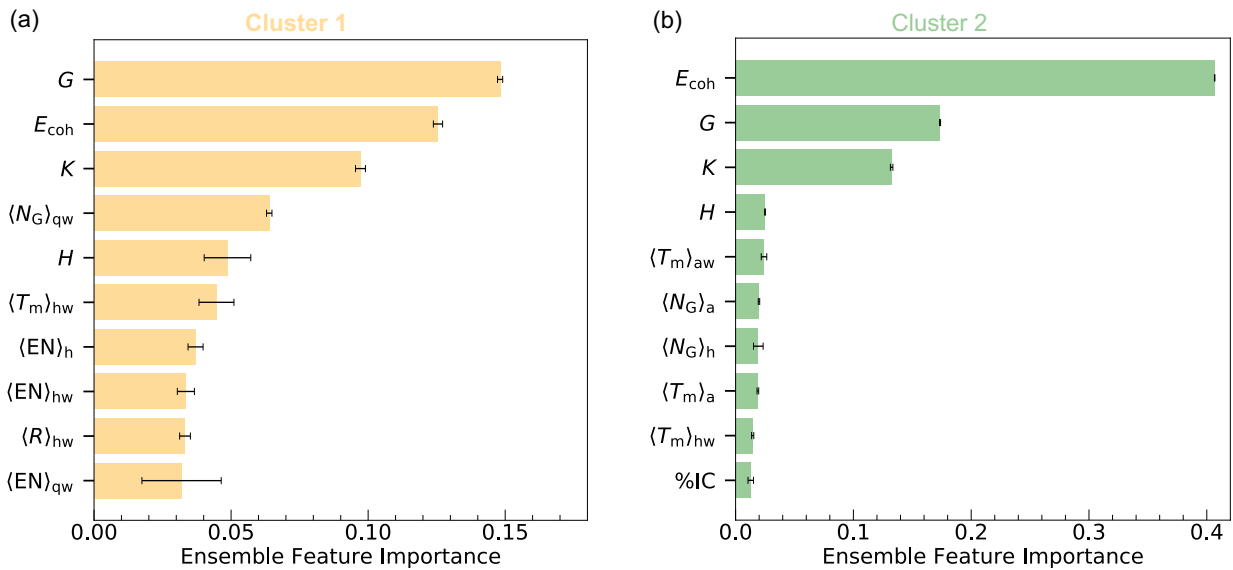

**Figure S10.** Feature importances from an ensemble tree method analysis for (a) cluster 1 and (b) cluster 2. Features are ranked in order of their importance from top to bottom. Only the top ten most important features are present for both clusters. The ensemble tree importance analysis represents ensemble-averaged results over ten seeds, and the horizontal error bars indicate values of importances one standard deviation above and below their respective average values over ten different seeds.

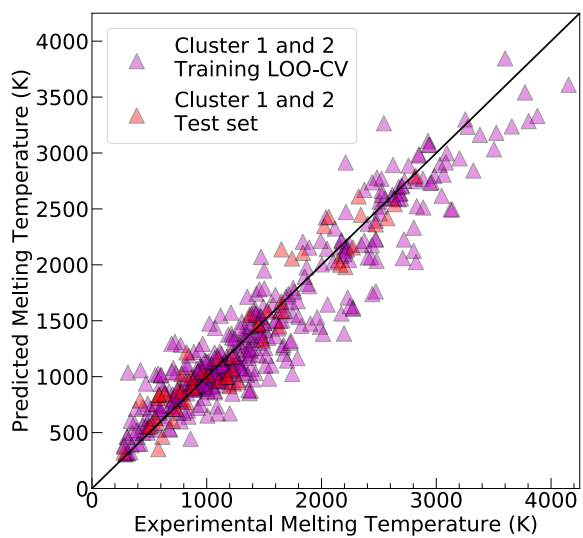

**Figure S11.** Correlation plot between the experimental and predicted melting temperatures for the best models trained on the training sets of clusters 1 and 2 and shown here in a combined way without specifying the cluster belongings. This plot is to help guide the comparison of the performance of the two-step approach to the one-step approach shown in **Table II** in the main manuscript. The training LOO-CV and test set labels represent ensemble averaged predictions over ten seeds. **Table III** (main manuscript) shows the reported error metrics. The random forest regression was the best model for clusters 1 and 2.

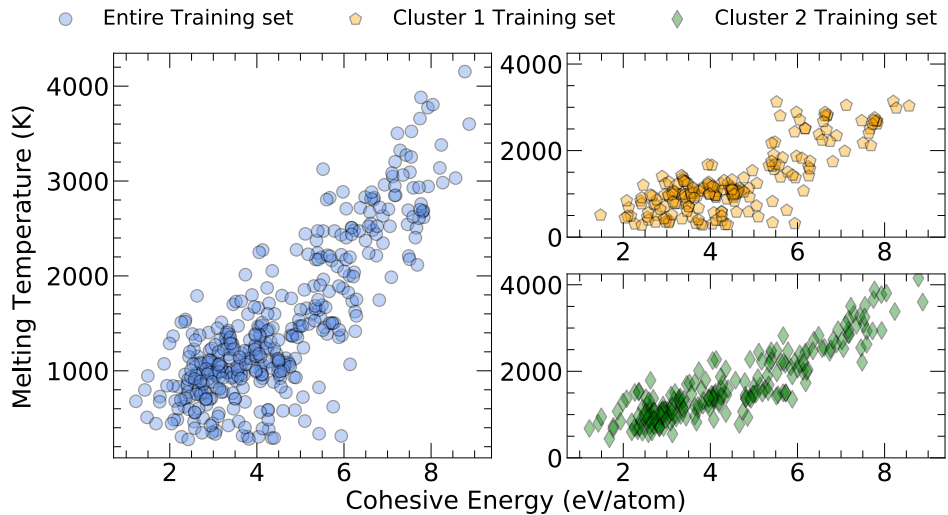

**Figure S12.** Correlation plots between the cohesive energy feature and the experimental melting temperatures for the entire training set and the training set of clusters 1 and 2. The Pearson correlation coefficients between the cohesive energy and melting temperature are 0.82, 0.80, and 0.90 for the training sets of the entire data set and clusters 1 and 2, respectively.

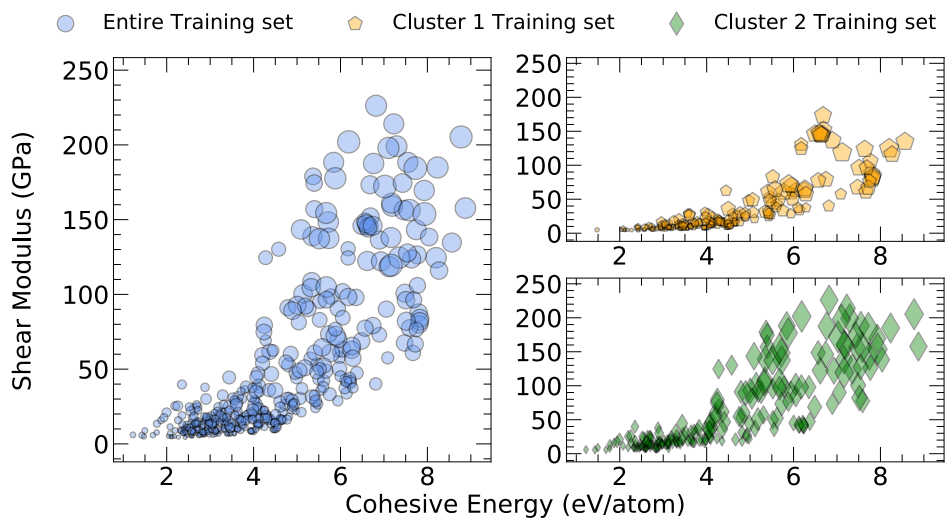

**Figure S13.** Correlation plots between the cohesive energy feature and the shear modulus feature for the entire training set and the training set of clusters 1 and 2. The marker size indicates the value of bulk modulus; the larger the marker size, the larger the bulk modulus value of the given material. The Pearson correlation coefficients between the cohesive energy and shear modulus are 0.79, 0.82, and 0.84 for the
8

training sets of the entire data set and clusters 1 and 2, respectively. The Pearson correlation coefficients between the cohesive energy and bulk modulus are 0.83, 0.86, and 0.86 for the training sets of the entire data set and clusters 1 and 2, respectively. The Pearson correlation coefficients between the shear and bulk moduli are 0.93, 0.90, and 0.94 for the training sets of the entire data set and clusters 1 and 2, respectively.

**Figure S14.** Elemental counts of the training sets of (a) cluster 1 and (b) cluster 2. These periodic tables show the elements found in the data set and, below the element symbol – the number of element occurrences. Elements not found in the binary compounds in the given data set are omitted from the periodic table figure for the respective set. Most occurring elements and their numbers of occurrence are colored in white.

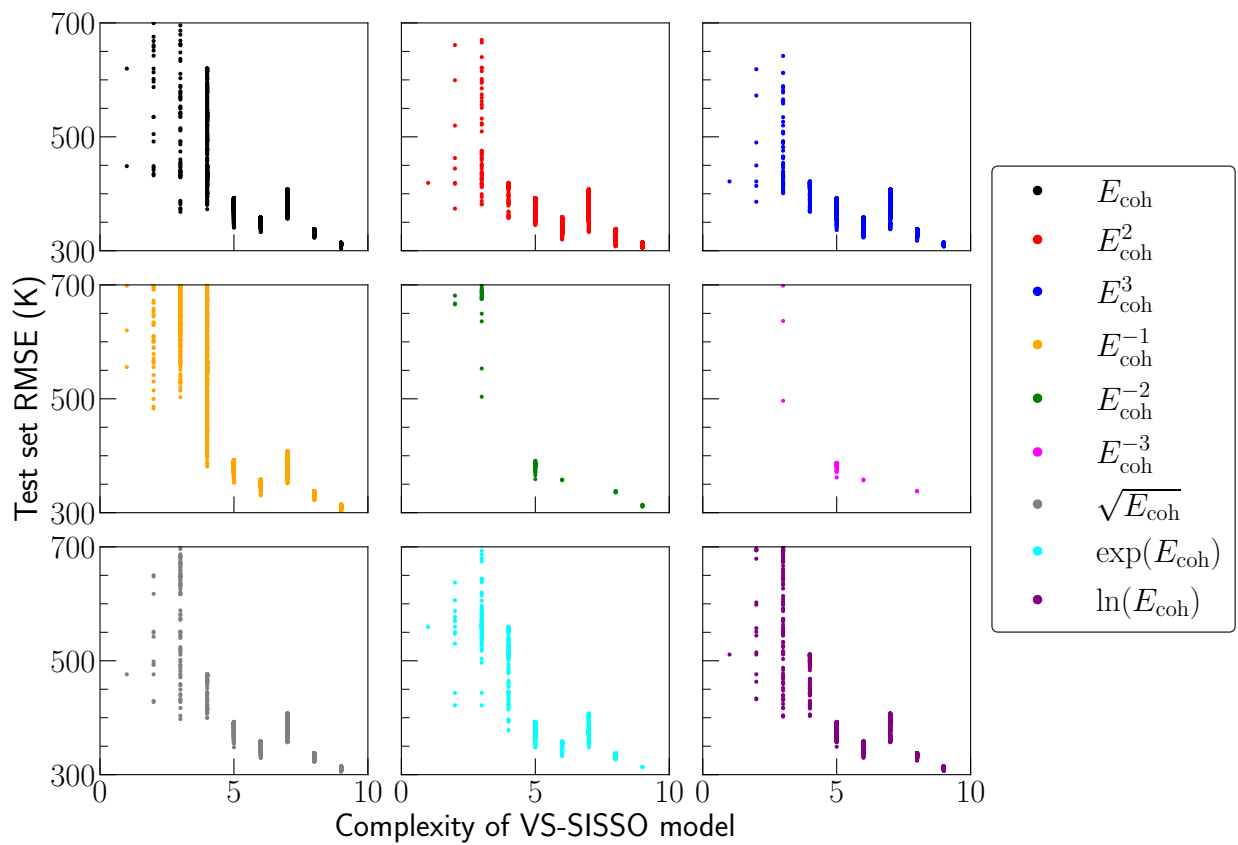

**Figure S15.** Operators acting on the cohesive energy feature versus the RMSE of the entire test set for VS-SISSO models. The RMSE scores are very similar, and thus, one best-performing operator can't be chosen. The $x$-axis shows the complexity of the VS-SISSO model defined by $3(D-1)+f$, where $D$ and $f$ are the dimensionality and feature complexity, respectively.

**Table S1.** Error metrics for each model type considered in this work and the best-performing combination of features used for the one-step approach and individual clusters in the two-step approach. The best-performing combination of features within each model type is chosen to minimize the average of the MAE and RMSE scores of the test set and the LOO-CV and 5-fold CV of the training set. For comparison purposes, models trained using standardized features as inputs and the logarithm of the melting temperatures as outputs are presented below (these models often resulted in the best performances). We also attempted model stacking but decided not to report more complex and less interpretable models with only slightly better performance here.

Abbreviations for the models used: LR = linear regression with or without L1 and/or L2 regularization, SVR = support vector regression, KRR = kernel ridge regression, GPR = Gaussian process regression, RFR = random forest regression, GBR = gradient-boosted tree regression.

Abbreviations for the input features used: C = all 6 compound features, and $\langle Z \rangle_{s/w}$ = concatenated composition-based and composition-agnostic weighting of the statistical combination s of all 8 elemental properties (refer to **Error! Reference source not found.** for naming conventions for the statistical combination subscript s). $\oplus$ Represents the concatenation operator. For all models, we considered fitting on (i) all 86 features, (iii) on compound features only (6 features), (iii) combined compound features with composition-based and composition-agnostic weighting for all unique statistical combinations of all elemental properties (22 features and 5 different fittings), and (iv) combined compound features with composition-based and composition-agnostic weighting for all unique pairs of statistical combinations of all elemental properties (38 features and 10 different fittings). Best models always included all six compound features, and excluding them resulted in the worst-performing models. Shown in bold are the best-performing models from **Error! Reference source not found.**.

| Approach | Cluster | Model | Features | Training set (LOO-CV) | | Training set (5-fold CV) | | Test set | |
|---|---|---|---|---|---|---|---|---|---|
| | | | | MAE [K] | RMSE [K] | MAE [K] | RMSE [K] | MAE [K] | RMSE [K] |
| One-step | - | LR | all | 255 | 345 | 269 | 354 | 270 | 345 |
| Two-step | 1 | LR | all | 220 | 329 | 234 | 338 | 223 | 306 |
| Two-step | 2 | LR | $C \oplus \langle Z \rangle_{g/w}$ | 265 | 352 | 280 | 378 | 235 | 273 |
| One-step | - | SVR | $C \oplus \langle Z \rangle_{g/w}$ | 187 | 273 | 207 | 283 | 273 | 360 |
| Two-step | 1 | SVR | $C \oplus \langle Z \rangle_{a/w} \oplus \langle Z \rangle_{h/w}$ | 201 | 299 | 201 | 303 | 174 | 252 |
| Two-step | 2 | SVR | $C \oplus \langle Z \rangle_{h/w} \oplus \langle Z \rangle_{g/w}$ | 197 | 278 | 226 | 315 | 214 | 278 |
| One-step | - | KRR | $C \oplus \langle Z \rangle_{g/w}$ | 199 | 285 | 209 | 285 | 223 | 306 |
| Two-step | 1 | KRR | $C \oplus \langle Z \rangle_{a/w} \oplus \langle Z \rangle_{s/w}$ | 194 | 285 | 210 | 321 | 207 | 286 |
| Two-step | 2 | KRR | $C \oplus \langle Z \rangle_{h/w} \oplus \langle Z \rangle_{q/w}$ | 224 | 295 | 247 | 324 | 214 | 255 |
| One-step | - | GPR | $C \oplus \langle Z \rangle_{a/w} \oplus \langle Z \rangle_{h/w}$ | 201 | 284 | 247 | 346 | 198 | 272 |
| Two-step | 1 | GPR | $C \oplus \langle Z \rangle_{h/w} \oplus \langle Z \rangle_{g/w}$ | 178 | 272 | 203 | 292 | 174 | 244 |
| Two-step | 2 | GPR | $C \oplus \langle Z \rangle_{g/w}$ | 214 | 287 | 269 | 369 | 236 | 286 |
| One-step | - | RFR | $C \oplus \langle Z \rangle_{h/w} \oplus \langle Z \rangle_{q/w}$ | 194 | 274 | 216 | 296 | 163 | 213 |
| **Two-step** | **1** | **RFR** | $\mathbf{C \oplus \langle Z \rangle_{h/w} \oplus \langle Z \rangle_{q/w}}$ | **169** | **246** | **189** | **264** | **136** | **169** |
| **Two-step** | **2** | **RFR** | $\mathbf{C \oplus \langle Z \rangle_{a/w} \oplus \langle Z \rangle_{h/w}}$ | **198** | **267** | **211** | **276** | **189** | **229** |
| **One-step** | **-** | **GBR** | $\mathbf{C \oplus \langle Z \rangle_{h/w} \oplus \langle Z \rangle_{q/w}}$ | **177** | **243** | **201** | **266** | **165** | **218** |
| Two-step | 1 | GBR | $C \oplus \langle Z \rangle_{a/w} \oplus \langle Z \rangle_{h/w}$ | 176 | 260 | 182 | 260 | 158 | 192 |
| Two-step | 2 | GBR | $C \oplus \langle Z \rangle_{g/w} \oplus \langle Z \rangle_{q/w}$ | 219 | 292 | 243 | 321 | 176 | 221 |